# Design and fabrication of guiding patterns for topography-based searching of 2D devices for scanning tunneling microscopy measurements

Huandong Chen[1,4], Hong Li[1,4], Yutao Li[1], He Zhao[1], Ming Lu[3], Kazuhiro Fujita[1*], Abhay N. Pasupathy[1,2*]

[1]Condensed Matter Physics and Materials Science Department, Brookhaven National Laboratory, Upton, NY, 11973, USA

[2]Department of Physics, Columbia University, New York, NY, 10027, USA

[3]Center for Functional Nanomaterials, Brookhaven National Laboratory, Upton, NY, 11973, USA

[4]These authors contributed equally: Huandong Chen and Hong Li

*Email: kfujita@bnl.gov, apn2108@columbia.edu

**Abstract**

We report the design and fabrication of guiding patterns for topography-based searching of two-dimensional (2D) devices for scanning tunneling microscopy (STM) measurements. Sub-micron geometric coordinate markers were etched into $SiO_2$/Si wafers, serving as both substrates for 2D device integration and guiding maps for sample navigation. Here, we used a monolayer graphene/h-BN device with an active area of smaller than 20 μm × 20 μm as a model system and demonstrated that the device could be reliably located in STM solely through topographic imaging of the guiding patterns and in situ stage calibration, without reliance on optical viewports or capacitive sensing. Atomically resolved topographic imaging and tunneling spectroscopy were also obtained. Our proposed navigation strategy is fully compatible with standard device fabrication procedures and requires no hardware modification to existing STM setups. As such, it offers a practical alternative for locating miniaturized devices in STM, shedding light on studying emerging quantum phenomena in 2D systems.

## Introduction

Scanning tunneling microscopy and spectroscopy (STM/S) have emerged as powerful experimental techniques for probing two-dimensional (2D) materials and their heterostructures[1-3] beyond the capabilities of conventional electronic transport and optical characterization. STM offers atomically resolved topographic imaging, enabling direct visualization of moiré patterns and other nanoscale structural features, while differential conductance ($dI/dV$) spectroscopy provides spatially resolved measurements of the local electronic density of states.[4-6] Recent breakthroughs include studies of superconductivity and correlated electronic phenomena in twist-engineered graphene systems,[7-10] as well as the observation of Wigner crystallization in transition metal dichalcogenide (TMD) heterostructures,[3,11] thereby driving a growing interest in exploring emerging quantum phases in 2D systems using STM.

Despite these advances, preparing 2D devices for STM measurements remains technically challenging.[12] A primary limitation lies in the difficulty of precisely locating microscale 2D devices within the STM field of view. Unlike measurements carried out on millimeter-sized bulk single crystals or wafer-scale thin films, where the STM tip can be reliably guided to the active region without much risk,[13-16] locating micro-scaled devices (typically 10-20 μm laterally) requires a designated searching process involving coarse xy-stage motion, which is critical for avoiding tip crashes and enabling meaningful data collection. One intuitive solution is to incorporate an optical viewport to visually guide the tip toward the device.[17-19] For instance, Zhao et al. employed a circular fan pattern to visually align the tip to active device region before inserting the STM module for cooling, which has proven working for graphene devices in a specially designed STM setup.[20,21] Most STM studies of 2D devices under extreme conditions[22-25] have relied on a capacitance-based navigation strategy developed in the early 2010s,[26] which uses a high-precision measurement setup to detect subtle capacitance changes as the tip scans across conductor-insulator boundaries.

Alternatively, topographic imaging, one of the core functionalities of STM, can, in principle, be utilized to navigate toward micro-scaled 2D devices. However, due to the inherently limited field of view of topographic scans (typically 1 – 1.5 µm laterally) and the incompatibility with common insulating substrates such as $SiO_2$ and h-BN, this approach has rarely been successfully implemented. In this work, by carefully design on-chip guiding patterns and tailoring the associated device fabrication procedures, we demonstrate a viable topography-based navigation strategy for locating 2D devices for STM measurements. Specifically, sub-micron angled geometric features are lithographically encoded in the guiding map to represent spatial coordinates (latitude and longitude), while standard bar patters are used to locally calibrate the coarse sample stage motions. Together, these practices allow the 2D device to be precisely located within the STM field of view. Importantly, our proposed strategy requires no hardware modifications or upgrades to the existing STM systems equipped with xy-stage motion sensors and is fully compatible with STM platforms operating under extreme temperatures and magnetic fields, thereby serving as an effective and complementary alternative to capacitance-based navigation techniques.

## Design principles of STM guiding patterns

We designed and implemented three different types of geometrical patterns to sequentially guide the STM tip toward the targeted 2D flake, as illustrated in Figure 1a. First, a set of coordinate-encoded circular sectors enables coarse localization of the sample area with an accuracy of approximately 100 μm × 100 μm. Second, linearly arranged half- and quarter-sector patterns mark the edges and corners of each square cell within the guiding map, allowing for deterministic and high-precision positioning. Third, a series of linearly arranged bar-shaped features near the center are employed to locally calibrate the stage motion along both the x- and y-axes, ensuring accurate tip landing during the final approach.

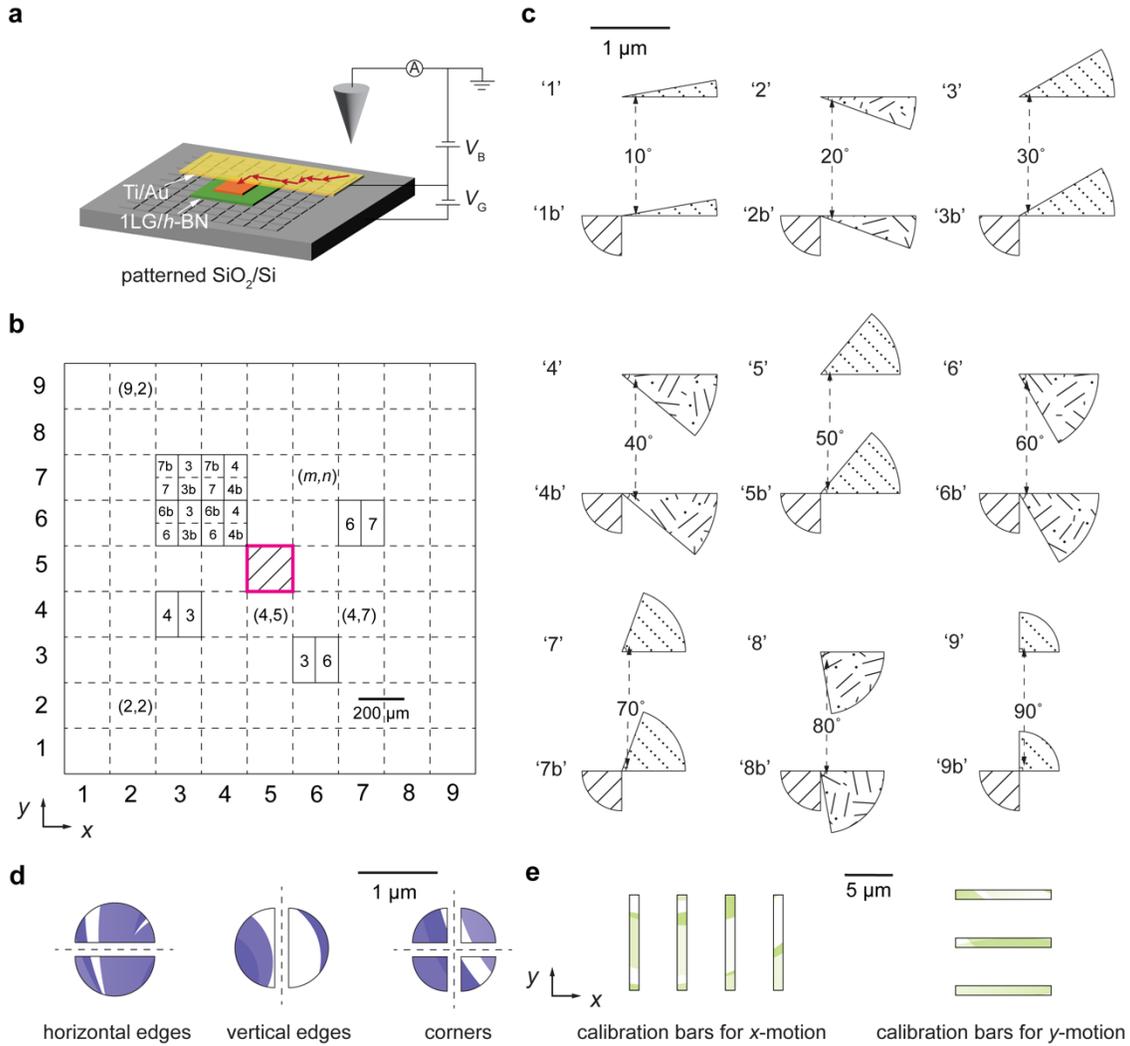

**Figure 1 Design principals of STM guiding patterns for topography-based navigation of 2D devices.** (**a**) Schematic illustration of a monolayer graphene/h-BN device integrated on a guiding chip for STM measurement. (**b**) Representative portions of the guiding map with encoded indices for all 100 μm × 100 μm subregions except the central four. (**c**) A full list of circular sector patterns used to encode numerical indices ranging from 1 to 9 and their corresponding b-variants. (**d**) Geometrical patterns employed to encode edges and corners of each 100 μm × 100 μm squared region. (**e**) Bar-shaped patterns used to locally calibrate x- and y-motions of the sample stage.

The entire active region of the guiding map spans 1.8 mm × 1.8 mm and is first divided into a 9 × 9 array of square cells, each measuring 200 μm × 200 μm, with central unpatterned area reserved for 2D device integration (Figure 1b). Each square is further subdivided into two rectangles (100 μm × 200 μm), with the left and right halves labeled with y- and x-coordinates, respectively, using numerical indices ranging from 1 to 9. These coordinates are encoded through circular sector patterns, with angular spans from 10° to 90° in 10° increments, positioned in either the first or fourth quadrant of the circle, as illustrated in Figure 1c. To distinguish between the upper and lower halves of each rectangular region, an additional set of indices, denoted from 1b to 9b, is defined by adding a quarter-sector feature in the third quadrant. Importantly, circular sectors representing odd numbers (1, 3, 5, 7, 9, and their corresponding b-variants) are consistently located in the first quadrant, whose those for even numbers appear in the fourth quadrant. The detailed spatial arrangement of these patterns is illustrated in Figure S1. This intentional angular partitioning significantly reduces the risk of misidentification due to imperfect angular resolution during topographic decoding.

Figure S2 presents the complete guiding map with encoded indices for all 100 μm × 100 μm subregions (excluding the central four), each populated with a specific geometric pattern that is periodically arranged with 1 μm pitch along the y-axis and 2 μm along the x-axis. This compact tiling ensures that any 1.5 μm × 1.5 μm topographic scan will capture at least part of a geometric motif, regardless of the STM tip's landing point. Furthermore, due to the specially structured indexing scheme, the landing position can be uniquely determined, albeit coarsely (with ~ 100 μm precision), from at most three consecutive topographic scans spaced 100 μm apart each.

Following the guiding map, the STM tip can be incrementally steered toward the vicinity of the unpatterned device region by translating the x-y sample stage several times in 100 μm steps.

Once near the target area, the boundary between adjacent squares, marked by linearly arranged half-circles (Figure 1d and Figure S3), can be identified by executing repeated forward-backward stage motions with progressively reduced step sizes. The precise location of a corner (~ 1 µm accuracy) is then determined through continuous topographic scanning along this boundary. At this stage, the relative coordinates of the device surface with respect to this corner position in real space can be obtained by correlating the STM-acquired location with high-resolution optical microscopic images taken during device fabrication. An alternative search strategy might involve first locating a known corner position upon initial tip approach and then navigating toward the central device region following the grids. However, this method often proves inefficient due to small rotational misalignments between the sample stage and the guiding substrate, which can arise during the sample mounting and loading processes.

The final step requires the STM tip to travel across a relatively long distance of several tens of microns to reach the small active device region (10-20 µm) from the nearest known position, while avoiding tip crash onto surrounding insulating surfaces such as h-BN and $SiO_2$. However, the stage displacement readout values may deviate from actual movements, owing to nonlinear piezoelectric responses and/or outdated calibration files. To address this, we implement a local calibration strategy using a set of bar-shaped patterns (1 µm wide, spaced 5 µm apart) precisely defined by electron-beam (E-beam) lithography, as illustrated in Figure 1e and Figure S4. Such patterns allow for direct calibration of x- and y-axis displacements and provides an accurate measure of any rotation misalignment between the substrate and the stage. These calibration results are then used to compensate for systematic deviation, thereby enabling highly controlled tip landing on the targeted device surface.

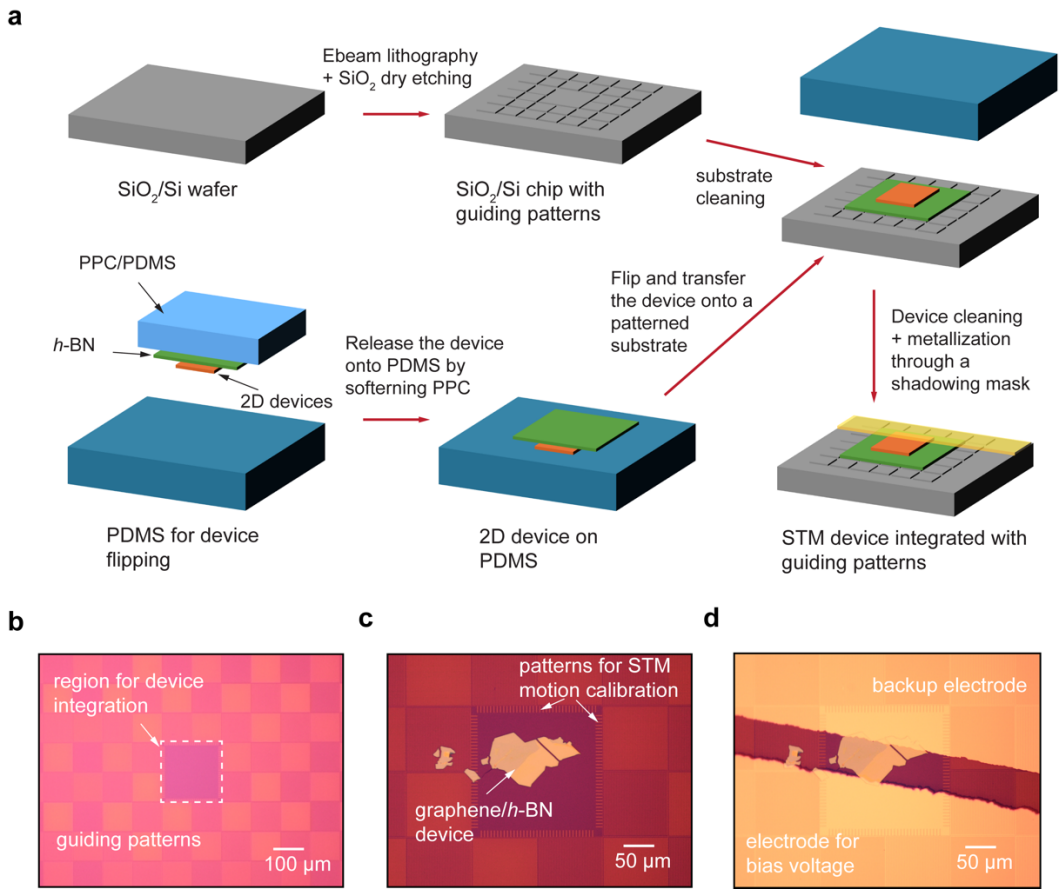

**Figure 2 Fabrication of STM guiding patterns and 2D device integration.** (**a**) Schematic illustration of fabrication process flow for guiding chips and the associated 2D device integration for STM measurements. (**b**) to (**d**) Optical micrographs of the STM guiding chip after RIE process (**b**), 2D device integration (**c**), and contact formation (**d**).

## Fabrication of STM guiding patterns and 2D device integration

The STM guiding patterns (1.8 mm × 1.8 mm) were fabricated on regular $SiO_2$/Si substrates combining E-beam lithography and $SiO_2$ reactive ion etching (RIE), as illustrated in the process flow in Figure 2a. A 100 kV E-beam lithography system was employed to write arrays of guiding patterns (5 × 5 per wafer) with sub-micron resolution on a 4-inch $SiO_2$/Si wafer, using polymethylmethacrylate (PMMA) as the resist. A relatively high beam current of 10 nA was used to accelerate the overall writing process (~ 15 min per chip), although the patterning resolution could be slightly compromised. Nonetheless, it did not affect the decoding of geometric shapes due to the robustness of the angle-based encoding scheme as described in the previous section. The patterned features were then transferred into the $SiO_2$ layer by $CHF_3/O_2$-based with an optimized etch depth of ~ 40 nm (Figure 2b). This depth was sufficient to clearly reveal the geometric patterns after subsequent metal deposition, while also minimizing the risk of STM tip crashing during topographic scans. Following RIE, the PMMA layer was removed in acetone, after which the wafer was diced into individual chips for 2D device integration. Prior to device fabrication, each guiding chip was additionally cleaned with $O_2$ plasma to ensure surface cleanliness and to facilitate the device integration.

A monolayer graphene device (~ 20 μm × 20 μm) was fabricated on a pre-patterned $SiO_2$ guiding chip to serve as a model 2D system for device-searching demonstration and STM measurements. The device fabrication process was adapted from established procedures with some modifications.[27-29] Specifically, a polycarbonate (PC) stamp was used for the initial assembly of the 2D heterostructures at ~ 110˚C, ensuring a high pick-up yield and good interface quality.[27] To facilitate device flipping, an additional polypropylene carbonate (PPC) stamp[28] in combination with a polydimethylsiloxane (PDMS) planar sheet[29] was employed. In this process, the PPC stamp was used to re-pick up the entire pre-assembled device stack from a temporary $SiO_2$ substrate and release it onto PDMS for the subsequent flipping and deterministic integration. Alternatively, a PPC stamp alone could be used for both the assembly and the release of the device stack onto PDMS. Finally, the device was aligned and deterministically transferred onto the central unpatterned region of the guiding chip, adjacent to the bar-shaped calibration patterns, by slowly releasing the PDMS, as shown in Figure 2c.

To prepare the device for STM measurements, a series of cleaning procedures, including organic solvent rinsing,[8,23] ultra-high vacuum (UHV) annealing,[11,30,31] and atomic force microscopy (AFM) tip cleaning,[32,33] could be carried out to minimize polymer residues on the graphene surface. Notably, unlike STM devices that rely on optical viewports or capacitive mapping for sample navigation, where the contacts to the device can be established using only narrow electrodes, our topography-based strategy further requires the guiding patterns themselves to be electrically conductive. To this end, two large-area electrodes were deposited through a thin-Si beam-based shadow mask (~ 60 μm-wide beam), ensuring conductive top surfaces while preserving the cleanliness of the device's surface, as shown in Figure 2d. One electrode, directly contacting the device, was used to apply the bias voltage during measurements, while the second electrode could be configured to a gate contact when needed (e.g., using h-BN as the gate dielectric). This electrode design significantly reduces the probabilities of STM tip crashes on insulating $SiO_2$ surfaces during the initial approach, considering the relative area ratio of conducting to insulating regions.

**Topography-based searching process for 2D devices**

Here, we showcase representative navigation procedures for locating a 2D device using the guiding patterns, as illustrated in Figure 3a. After the STM tip was approached the chip surface, an initial topographic scan of a 1.5 μm × 1.5 μm area was performed. The acquired feature, a circular sector in the first quadrant, was identified as the odd index '7' based on its opening angle, despite the non-ideal resolution of the fabricated patterns (Figure 3b, left). The presence of an additional quarter sector in the third quadrant further confirmed the designation as '7b', rather than '7'. At this stage, the decoded pattern corresponded to 18 possible locations on the guiding map, as illustrated in Figure S5. To further refine the location, the tip was displaced ~ 100 μm to the right of the original landing place by coarsely driving the sample stage toward the opposite direction, guided by the default x-y sensor readout. The subsequent scan was decoded as '4' from the scan (Figure 3b, middle), reducing the possible positions to only two. A further ~100 μm downward displacement decoded the pattern as '4b' from the topographic scan (Figure 3b, right), which uniquely determined the tip location within a specific 100 μm × 100 μm square. It is worth noting that the specific arrangement of this guiding map ensures the coarse determination of the absolute tip position by decoding any three neighboring patterns.

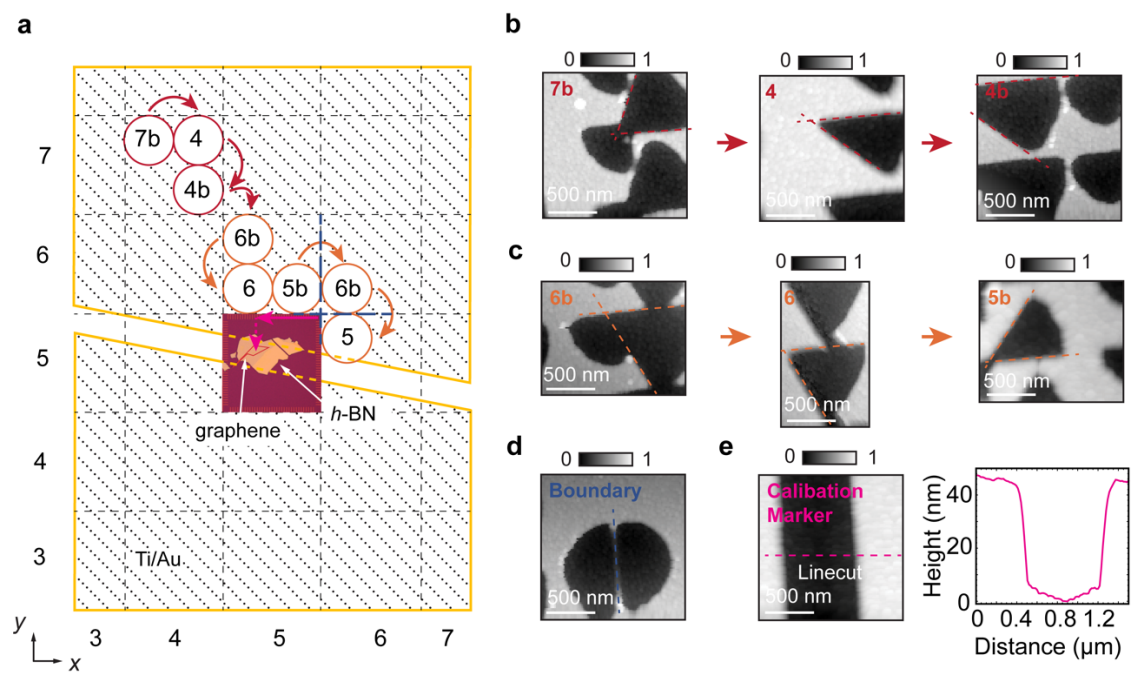

**Figure 3 Representative topography-based navigation process for a 2D device.** (**a**) Schematic illustration of the overall topography-based navigation process for a monolayer graphene/h-BN device presented on the guiding map. (**b**) STM topographic scans of numerically encoded patterns showing the process of coarse position determination. (**c**) Topographic images of numerically encoded patterns showing the process of incrementally navigating toward the device region. (**d**) Topographic image of a boundary marker used for precise position determination. (**e**) Topographic image of a bar-shaped calibration marker for stage motion along the x-axis (left) and the corresponding height profile of the linecut (right).

With the rough position established, the tip was then navigated toward the unpatterned device region either incrementally, e.g., following a path from '6b' to '6' to '5b' (Figure 3c), or directly by moving 200 µm rightward and 300 µm downward. The vertical boundary between squares '5b' (left) and '6b' (right) was then identified by successive scans with decreasing step sizes through a few attempts, as shown in Figure 3d. Following similar procedures, the horizontal boundary between squares '6b' (top) and '5' (bottom) was located, thereby defining the top-right corner of the device region with high precision. To further ensure accurate landing, the stage displacement readout in both x- and y-directions was calibrated locally using a set of bar-shaped fiducial patterns (10 µm × 1 µm) separated by 5 µm (Figure 3e). This calibration revealed offsets of ~14% in the x-direction and ~10% along the y-direction, which were subsequently corrected. In addition, a small rotation misalignment (< 4°), attributed to sample mounting process, was also measured and compensated. These local calibrations are essential for safely and accurately landing the STM tip on microscale active device regions, particularly in systems with small active areas such as TMD devices (< 10 µm), without crashing onto adjacent insulating surfaces.

Thus far, the entire navigation process typically takes from half a day to a day, which is still a reasonable time frame for high-quality STM measurements that often last for several days or weeks. It is also worth emphasizing that x- and y-motion calibrations are not generally necessary during routine operations unless the device is placed far from the stage center or possesses an exceptionally small active area. Furthermore, in most cases a full topographic image is not required to decode the guiding pattern; partial scans containing angle information are often sufficient, thereby further accelerating the overall device-searching process.

## STM measurements of a monolayer graphene/h-BN device

All STM measurements presented in this study were conducted at a base temperature of 4.5 K and in zero magnetic field, serving as a proof-of-concept demonstration, although the instrument is capable of operating down to 300 mK and under out-of-plane magnetic field up to 11 T. A monolayer graphene device was selected as the model 2D system for device navigation experiments due to its air-stability,[34] ease of fabrication,[35] and the relative simplicity of surface cleaning.[33]

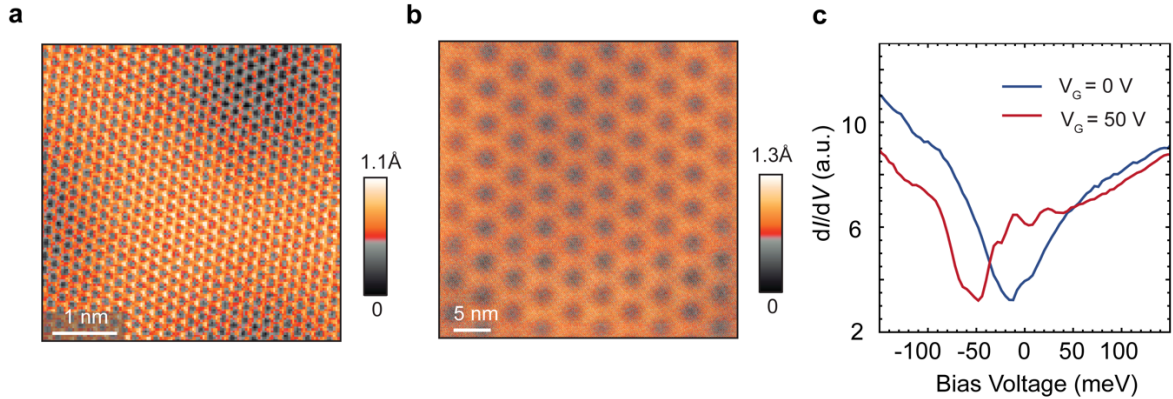

**Figure 4 STM/S measurements of a monolayer graphene/h-BN device.** (**a**) STM topographic image of the device with a 5 nm field of view. (**b**) 40 nm × 40 nm topographic scan of the graphene surface showing the absence of surface contaminations and a moiré pattern formed from monolayer graphene/h-BN heterostructure. (**c**) d$I$/d$V$ spectra of monolayer graphene at back gate voltages of 0 V (blue) and 50 V (red).

Upon landing on the graphene surface, high-resolution topographic imaging with a 5 nm field-of-view (Figure 4a) resolved the honeycomb lattice at the atomic scale. A representative 40 nm × 40 nm scan of a clean graphene region (Figure 4b) revealed not only the absence of surface contamination but also a well-defined hexagonally arranged superlattice with a periodicity of 0.2 nm. The moiré pattern observed here is attributed to the heterostructure consisting of the monolayer graphene and the underlying h-BN, with a twist angle of approximately 2.3° that was unintentionally introduced during the stacking process. Figure 4c plots d$I$/d$V$ spectra of monolayer graphene at 0 V and 50 V gate voltages applied through the $SiO_2$ dielectric, revealing a clear spectrum shift upon applying gate voltages.

These high-quality STM/S data shown here have clearly indicated the success of this topography-based searching strategy, which is broadly applicable to other 2D materials and heterostructures, including TMDs, provided that pristine surfaces are preserved during device assembly and flipping. Moreover, the strategy readily accommodates advanced STM measurements such as density-of-states mapping as functions of temperature and external magnetic field, thereby enabling comprehensive exploration of emergent phenomena in 2D systems.

**Conclusion and discussion**

In conclusion, we have successfully demonstrated a topography-based strategy for locating microscale 2D devices in STM, enabled by the design and fabrication of coordinate-encoded guiding patterns. The integration of 2D devices on $SiO_2$-based guiding chips is fully compatible with standard device fabrication workflows. Compared with navigating methods relying on optical viewports and visual alignment,[11,17,20] our strategy provides the possibility of deterministically locating the microscale devices, and hence is promising in handling TMD devices or twist-engineered heterostructures with sub-10 μm active regions. Moreover, unlike the capacitance mapping method that requires a capacitance bridge setup or a back-gate device configuration to detect the subtle capacitance change between the tip and sample (~ $10^{-18}$ F) across the conductor-insulator interface,[22,23,26] our technique only uses topographic scanning, one of the most basic functions of STM, for navigation. Therefore, our strategy is generally applicable to any existing STM setup with xy-stage motions, requiring no additional electronics or special device configurations. Overall, our proposed approach provides a practical, or even advantageous, alternative to capacitance-based mapping for STM systems without optical viewports, thereby offering more possibilities for studying versatile 2D devices using STM, particularly under cryogenic temperatures and high magnetic fields.

Importantly, the navigation protocol strictly follows a decision-tree logic that subsequently guides the STM tip toward the active device area. This deterministic framework naturally lends itself to future automation through pattern-recognition and machine-learning algorithms, which could significantly reduce searching time and further enhance accuracy. Moreover, the device fabrication and navigation strategies discussed here can be readily extended beyond 2D materials to accommodate cleaved single crystal with intriguing physical properties yet limited dimensions, thereby broadening the scope of STM studies across a wide range of emerging quantum systems.

## Methods

**Fabrication of $SiO_2$-based guiding patterns**

The sub-micron featured guiding patterns were prepared on regular $SiO_2$/Si substrates (285 nm thick thermal oxide, Nova) using a 100-kV Ebeam lithography system (JEOL JBX-6300FS) and PMMA as the resist. The patterns were transferred onto $SiO_2$ by reactive ion etching using $CHF_3$ and $O_2$ gases to achieve an etch depth of ~ 40 nm. PMMA was then stripped in acetone for 5 min and the patterned wafers were cleaned in IPA and DI subsequently.

**2D device assembly and integration for STM**

Graphene and h-BN flakes were freshly cleaved onto $SiO_2$ substrates ($O_2$ plasma treated, 2 min) at 100˚C, following a hot cleavage method.[36] A PC stamp was used to subsequently pick up an h-BN flake and a monolayer graphene at 110˚C, and then the entire stack was dropped onto a pristine $SiO_2$ substrate by melting PC film at 180˚C. After dissolving PC film in chloroform for 10 min, the device stack was re-picked up by a PPC stamp and dropped onto a planar PDMS ($O_2$ plasma treated, 2 min) for flipping at 110˚C.

The flipped device was aligned and transferred onto the centered region (200 μm × 200 μm) of a guiding-pattern-encoded $SiO_2$ substrate at 80˚C. The surface of graphene was cleaned by NMP-based organic solvents and an AFM tip in a contact mode (FX40, Park Systems), followed by metal deposition of Ti/Au=5/80 nm by Ebeam evaporation through a thin Si-based shadow mask. The shadow mask was fabricated from a 100 μm-thick Si wafer by cryogenic Si deep etch using 100 nm Ni film as the etch mask.

**STM/S measurements**

STM/S measurements were carried out using a Pt/Ir tip in a customized Unisoku USM1300 system at ~ 4.5 K and under zero magnetic field. 1.5 μm × 1.5 μm topographic images of guiding patterns were taken to determine the pattern locations during the device navigation process. Small-area atomically resolved topographic images of graphene and graphene/h-BN Moiré patterns were recorded at $V_{sample}$ = 10 mV and $I_{set}$ = 30 pA, where $V_{sample}$ is the setup bias voltage and $I_{set}$ is the setup current. d$I$/d$V$ spectra measurements were performed under the following conditions: $V_{sample}$ = 200 mV, $I_{set}$ = 400 pA, $V_{exc}$ = 4 mV, and 917 Hz frequency. $V_{exc}$ is the bias modulation generated by the lock-in amplifier. The back gate voltages were applied through the 285 nm thick $SiO_2$ dielectric.

# Author declarations

## Author contributions

H.C. conceived the idea and designed the experiments. A.N.P. and K.F. supervised the project. H.C. prepared guiding patterns and fabricated 2D devices for STM measurements, with input from Y.L. and M.L. H.L., H.Z., and H.C. carried out STM and AFM measurements. H.C., K.F., and A.N.P. wrote the manuscript with input from all other coauthors.

## Acknowledgements

The work at Brookhaven National Laboratory (BNL) is supported by the Office of Basic Energy Science, Materials Science and Engineering Division, U.S. Department of Energy (DOE) under contract No. DE-SC0012704. This research used Nanofabrication facilities of the Center for Functional Nanomaterials (CFN), which is a U.S. DOE Office of Science Facility, at BNL under contract No. DE-SC0012704.

## Data availability

The data are available from the corresponding author of the article on reasonable request.

## Conflict of interest

The authors declare no competing financial interests.

# Supplemental Information for

# Design and fabrication of guiding patterns for topography-based searching of 2D devices for scanning tunneling microscopy measurements

Huandong Chen[1,4], Hong Li[1,4], Yutao Li[1], He Zhao[1], Ming Lu[3], Kazuhiro Fujita[1*], Abhay N. Pasupathy[1,2*]

[1]Condensed Matter Physics and Materials Science Department, Brookhaven National Laboratory, Upton, NY, 11973, USA

[2]Department of Physics, Columbia University, New York, NY, 10027, USA

[3]Center for Functional Nanomaterials, Brookhaven National Laboratory, Upton, NY, 11973, USA

[4]These authors contributed equally: Huandong Chen and Hong Li

*Email: kfujita@bnl.gov, apn2108@columbia.edu

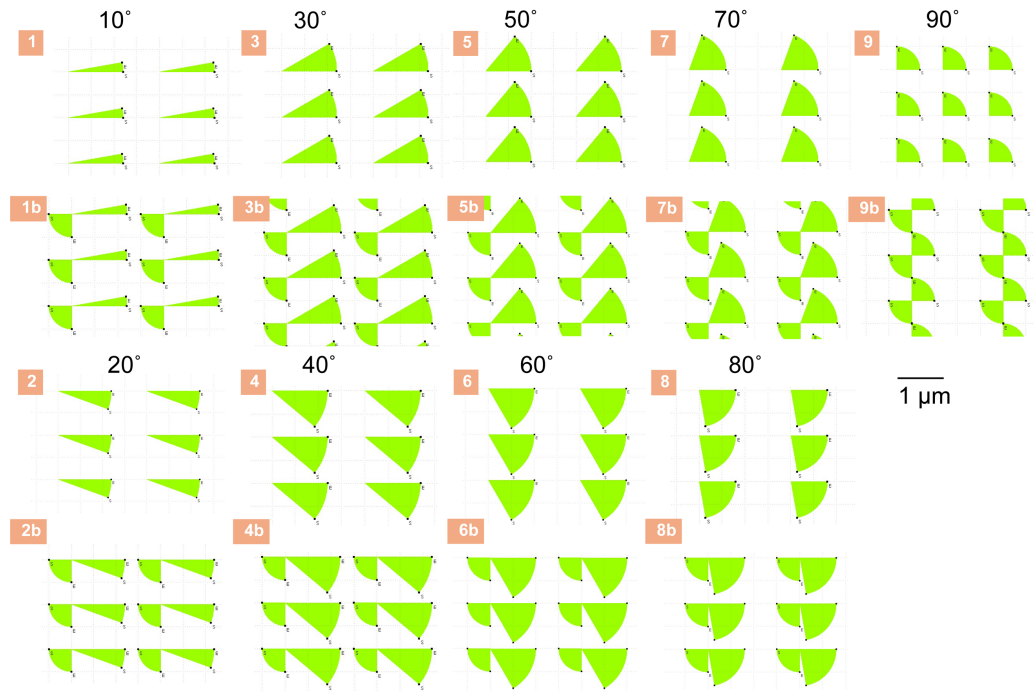

**Figure S1** Spatial arrangement illustration of circular sector patterns used to encode numerical indices from 1 to 9 and their corresponding b-variants.

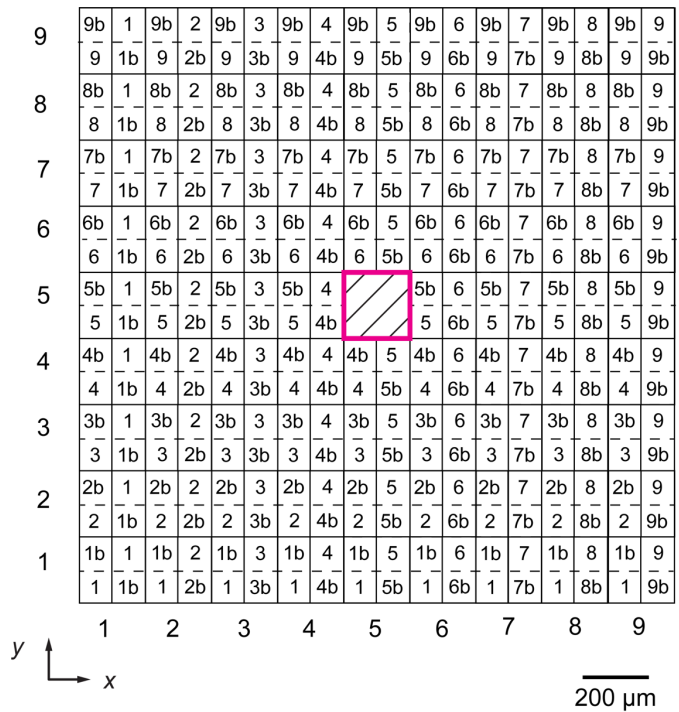

**Figure S2** Complete guiding map with encoded indices for all 100 μm × 100 μm subregions.

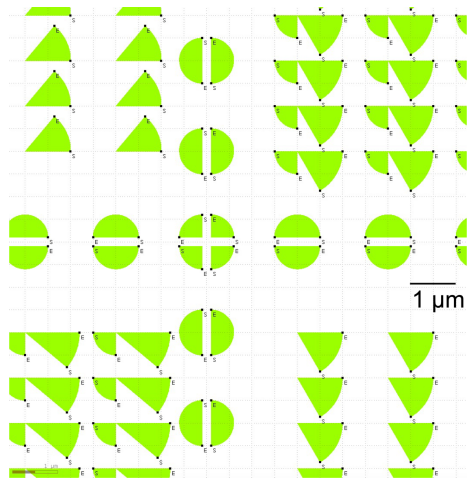

**Figure S3** Spatial arrangement illustration of horizontal and vertical edge patterns, as well as the corners at the boundaries of each 100 μm × 100 μm subregions.

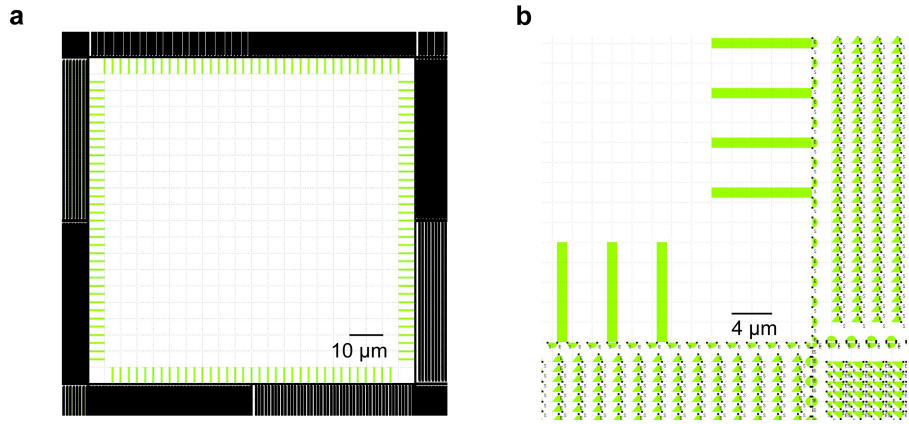

**Figure S4** Spatial arrangement illustration of device integration region (**a**) and bar-shaped calibration patterns for sample stage motions (**b**).

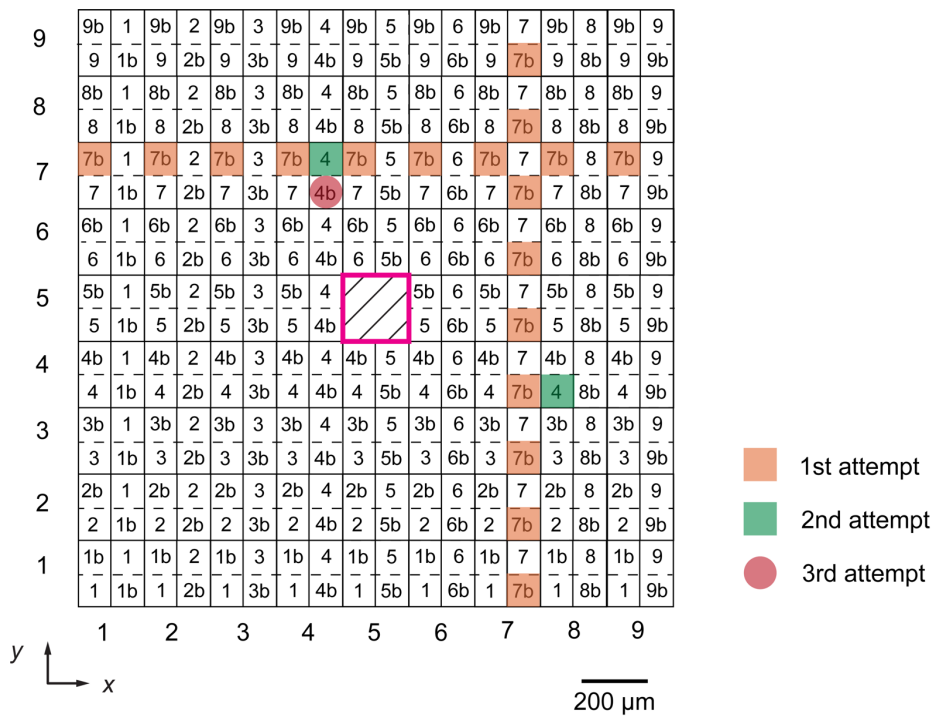

**Figure S5** Schematic illustration of the coarse location determination process. Three consecutive topographic scans spaced 100 μm apart can be used to unambiguously determine the tip location, with a precision of ~ 100 μm. The number of possible locations is reduced from 18 to 2, and then to 1 after each attempt.